\documentclass[aps,prl,twocolumn,superscriptaddress,showpacs,floatfix]{revtex4} 

\newcommand{\figsize}{0.5\textwidth}      
\usepackage{graphicx}
\usepackage{dcolumn}
\usepackage{bm}
\usepackage{setspace}
\usepackage{xspace}
\usepackage{amssymb}
\usepackage{color}
\usepackage{multirow}

\newcommand{\pt}{{p}_{\scriptscriptstyle\rm T}}
\newcommand{\et}{{E}_{\scriptscriptstyle\rm T}}
\newcommand{\met}{\mbox{$\protect \raisebox{.3ex}{$\not$}\et$}}
\newcommand{\etal}{{\em et al.}}

\newcommand{\wenu}{\ensuremath{W \rightarrow e \nu}\xspace}
\newcommand{\zee}{\ensuremath{Z \rightarrow e^{+}e^{-}}\xspace}
\newcommand{\zeenc}{\ensuremath{Z \rightarrow e e}\xspace}
\newcommand{\ztau}{\ensuremath{Z \rightarrow \tau^{+}\tau^{-}}\xspace}
\newcommand{\wtnu}{\ensuremath{W \rightarrow \tau \nu}\xspace}
\newcommand{\ppbar}{p\bar{p}}
\newcommand{\yW}{y_{W}}

\newcommand{\ifb}{\ensuremath{\rm fb^{-1}}\xspace}
\newcommand{\TeV}{\ensuremath{\mathrm{Te\kern-0.1em V}}\xspace}
\newcommand{\GeV}{\ensuremath{\mathrm{Ge\kern-0.1em V}}\xspace}
\newcommand{\metvec}{\mbox{${\not\!\! \vec{E}_T}$}}

\begin{document}

\title{Direct Measurement of the $W$ Production Charge Asymmetry in $\ppbar$ Collisions at $\sqrt{s} = 1.96$ TeV}
\affiliation{Institute of Physics, Academia Sinica, Taipei, Taiwan 11529, Republic of China} 
\affiliation{Argonne National Laboratory, Argonne, Illinois 60439} 
\affiliation{University of Athens, 157 71 Athens, Greece} 
\affiliation{Institut de Fisica d'Altes Energies, Universitat Autonoma de Barcelona, E-08193, Bellaterra (Barcelona), Spain} 
\affiliation{Baylor University, Waco, Texas  76798} 
\affiliation{Istituto Nazionale di Fisica Nucleare Bologna, $^x$University of Bologna, I-40127 Bologna, Italy} 
\affiliation{Brandeis University, Waltham, Massachusetts 02254} 
\affiliation{University of California, Davis, Davis, California  95616} 
\affiliation{University of California, Los Angeles, Los Angeles, California  90024} 
\affiliation{University of California, San Diego, La Jolla, California  92093} 
\affiliation{University of California, Santa Barbara, Santa Barbara, California 93106} 
\affiliation{Instituto de Fisica de Cantabria, CSIC-University of Cantabria, 39005 Santander, Spain} 
\affiliation{Carnegie Mellon University, Pittsburgh, PA  15213} 
\affiliation{Enrico Fermi Institute, University of Chicago, Chicago, Illinois 60637}
\affiliation{Comenius University, 842 48 Bratislava, Slovakia; Institute of Experimental Physics, 040 01 Kosice, Slovakia} 
\affiliation{Joint Institute for Nuclear Research, RU-141980 Dubna, Russia} 
\affiliation{Duke University, Durham, North Carolina  27708} 
\affiliation{Fermi National Accelerator Laboratory, Batavia, Illinois 60510} 
\affiliation{University of Florida, Gainesville, Florida  32611} 
\affiliation{Laboratori Nazionali di Frascati, Istituto Nazionale di Fisica Nucleare, I-00044 Frascati, Italy} 
\affiliation{University of Geneva, CH-1211 Geneva 4, Switzerland} 
\affiliation{Glasgow University, Glasgow G12 8QQ, United Kingdom} 
\affiliation{Harvard University, Cambridge, Massachusetts 02138} 
\affiliation{Division of High Energy Physics, Department of Physics, University of Helsinki and Helsinki Institute of Physics, FIN-00014, Helsinki, Finland} 
\affiliation{University of Illinois, Urbana, Illinois 61801} 
\affiliation{The Johns Hopkins University, Baltimore, Maryland 21218} 
\affiliation{Institut f\"{u}r Experimentelle Kernphysik, Universit\"{a}t Karlsruhe, 76128 Karlsruhe, Germany} 
\affiliation{Center for High Energy Physics: Kyungpook National University, Daegu 702-701, Korea; Seoul National University, Seoul 151-742, Korea; Sungkyunkwan University, Suwon 440-746, Korea; Korea Institute of Science and Technology Information, Daejeon, 305-806, Korea; Chonnam National University, Gwangju, 500-757, Korea} 
\affiliation{Ernest Orlando Lawrence Berkeley National Laboratory, Berkeley, California 94720} 
\affiliation{University of Liverpool, Liverpool L69 7ZE, United Kingdom} 
\affiliation{University College London, London WC1E 6BT, United Kingdom} 
\affiliation{Centro de Investigaciones Energeticas Medioambientales y Tecnologicas, E-28040 Madrid, Spain} 
\affiliation{Massachusetts Institute of Technology, Cambridge, Massachusetts  02139} 
\affiliation{Institute of Particle Physics: McGill University, Montr\'{e}al, Qu\'{e}bec, Canada H3A~2T8; Simon Fraser University, Burnaby, British Columbia, Canada V5A~1S6; University of Toronto, Toronto, Ontario, Canada M5S~1A7; and TRIUMF, Vancouver, British Columbia, Canada V6T~2A3} 
\affiliation{University of Michigan, Ann Arbor, Michigan 48109} 
\affiliation{Michigan State University, East Lansing, Michigan  48824}
\affiliation{Institution for Theoretical and Experimental Physics, ITEP, Moscow 117259, Russia} 
\affiliation{University of New Mexico, Albuquerque, New Mexico 87131} 
\affiliation{Northwestern University, Evanston, Illinois  60208} 
\affiliation{The Ohio State University, Columbus, Ohio  43210} 
\affiliation{Okayama University, Okayama 700-8530, Japan} 
\affiliation{Osaka City University, Osaka 588, Japan} 
\affiliation{University of Oxford, Oxford OX1 3RH, United Kingdom} 
\affiliation{Istituto Nazionale di Fisica Nucleare, Sezione di Padova-Trento, $^y$University of Padova, I-35131 Padova, Italy} 
\affiliation{LPNHE, Universite Pierre et Marie Curie/IN2P3-CNRS, UMR7585, Paris, F-75252 France} 
\affiliation{University of Pennsylvania, Philadelphia, Pennsylvania 19104}
\affiliation{Istituto Nazionale di Fisica Nucleare Pisa, $^z$University of Pisa, $^{aa}$University of Siena and $^{bb}$Scuola Normale Superiore, I-56127 Pisa, Italy} 
\affiliation{University of Pittsburgh, Pittsburgh, Pennsylvania 15260} 
\affiliation{Purdue University, West Lafayette, Indiana 47907} 
\affiliation{University of Rochester, Rochester, New York 14627} 
\affiliation{The Rockefeller University, New York, New York 10021} 
\affiliation{Istituto Nazionale di Fisica Nucleare, Sezione di Roma 1, $^{cc}$Sapienza Universit\`{a} di Roma, I-00185 Roma, Italy} 

\affiliation{Rutgers University, Piscataway, New Jersey 08855} 
\affiliation{Texas A\&M University, College Station, Texas 77843} 
\affiliation{Istituto Nazionale di Fisica Nucleare Trieste/Udine, I-34100 Trieste, $^{dd}$University of Trieste/Udine, I-33100 Udine, Italy} 
\affiliation{University of Tsukuba, Tsukuba, Ibaraki 305, Japan} 
\affiliation{Tufts University, Medford, Massachusetts 02155} 
\affiliation{Waseda University, Tokyo 169, Japan} 
\affiliation{Wayne State University, Detroit, Michigan  48201} 
\affiliation{University of Wisconsin, Madison, Wisconsin 53706} 
\affiliation{Yale University, New Haven, Connecticut 06520} 
\author{T.~Aaltonen}
\affiliation{Division of High Energy Physics, Department of Physics, University of Helsinki and Helsinki Institute of Physics, FIN-00014, Helsinki, Finland}
\author{J.~Adelman}
\affiliation{Enrico Fermi Institute, University of Chicago, Chicago, Illinois 60637}
\author{T.~Akimoto}
\affiliation{University of Tsukuba, Tsukuba, Ibaraki 305, Japan}
\author{B.~\'{A}lvarez~Gonz\'{a}lez$^s$}
\affiliation{Instituto de Fisica de Cantabria, CSIC-University of Cantabria, 39005 Santander, Spain}
\author{S.~Amerio$^y$}
\affiliation{Istituto Nazionale di Fisica Nucleare, Sezione di Padova-Trento, $^y$University of Padova, I-35131 Padova, Italy} 

\author{D.~Amidei}
\affiliation{University of Michigan, Ann Arbor, Michigan 48109}
\author{A.~Anastassov}
\affiliation{Northwestern University, Evanston, Illinois  60208}
\author{A.~Annovi}
\affiliation{Laboratori Nazionali di Frascati, Istituto Nazionale di Fisica Nucleare, I-00044 Frascati, Italy}
\author{J.~Antos}
\affiliation{Comenius University, 842 48 Bratislava, Slovakia; Institute of Experimental Physics, 040 01 Kosice, Slovakia}
\author{G.~Apollinari}
\affiliation{Fermi National Accelerator Laboratory, Batavia, Illinois 60510}
% ...

\collaboration{CDF Collaboration\footnote{With visitors from $^a$University of Massachusetts Amherst, Amherst, Massachusetts 01003,
$^b$Universiteit Antwerpen, B-2610 Antwerp, Belgium, 
$^c$University of Bristol, Bristol BS8 1TL, United Kingdom,
$^d$Chinese Academy of Sciences, Beijing 100864, China, 
$^e$Istituto Nazionale di Fisica Nucleare, Sezione di Cagliari, 09042 Monserrato (Cagliari), Italy,
$^f$University of California Irvine, Irvine, CA  92697, 
$^g$University of California Santa Cruz, Santa Cruz, CA  95064, 
$^h$Cornell University, Ithaca, NY  14853, 
$^i$University of Cyprus, Nicosia CY-1678, Cyprus, 
$^j$University College Dublin, Dublin 4, Ireland,
$^k$University of Edinburgh, Edinburgh EH9 3JZ, United Kingdom, 
$^l$University of Fukui, Fukui City, Fukui Prefecture, Japan 910-0017
$^m$Kinki University, Higashi-Osaka City, Japan 577-8502
$^n$Universidad Iberoamericana, Mexico D.F., Mexico,
$^o$Queen Mary, University of London, London, E1 4NS, England,
$^p$University of Manchester, Manchester M13 9PL, England, 
$^q$Nagasaki Institute of Applied Science, Nagasaki, Japan, 
$^r$University of Notre Dame, Notre Dame, IN 46556,
$^s$University de Oviedo, E-33007 Oviedo, Spain, 
$^t$Texas Tech University, Lubbock, TX  79609, 
$^u$IFIC(CSIC-Universitat de Valencia), 46071 Valencia, Spain,
$^v$University of Virginia, Charlottesville, VA  22904,
$^w$Bergische Universit\"at Wuppertal, 42097 Wuppertal, Germany,
$^{ee}$On leave from J.~Stefan Institute, Ljubljana, Slovenia, 
}}
\noaffiliation

\begin{abstract}
We present the first direct measurement of the $W$ production charge
asymmetry as a function of the $W$ boson rapidity $\yW$ in $\ppbar$
collisions at $\sqrt{s} = 1.96$ $\TeV$.  We use a sample of $\wenu$
events in data from 1~$\ifb$ of integrated luminosity collected using 
the CDF II detector. In the region $|\yW| <$ 3.0, this measurement is 
capable of constraining the ratio of up- and down-quark momentum 
distributions in the proton more directly than in previous measurements 
of the asymmetry that are functions of the charged-lepton pseudorapidity.
\end{abstract}


\maketitle
At the Fermilab Tevatron, where $\ppbar$ collisions are produced at
$\sqrt{s} = 1.96$ $\TeV$, $W^{+}(W^{-})$ bosons are created primarily
by the interaction of $u ~(d)$ quarks from the proton and $\bar{d}
~(\bar{u})$ quarks from the anti-proton. Since $u$ quarks carry, on
average, a higher fraction of the proton's momentum than $d$
quarks~\cite{mrst-pdf,cteq-pdf}, the $W^{+}$ tends to be boosted
along the proton beam direction and the $W^{-}$ tends to be boosted
along the anti-proton direction. The difference between the $W^{+}$
and $W^{-}$ rapidity distributions results in a charge asymmetry
\begin{equation}
A(y_W)  =  \frac{d\sigma^{+}/dy_{W} - d\sigma^{-}/dy_{W}}{d\sigma^{+}/dy_{W} + d\sigma^{-}/dy_{W}},
\label{eq:asym}
\end{equation}
where $\yW$ is the $W$ boson rapidity~\cite{coord} and 
$d\sigma^{\pm}/dy_{W}$ is the differential cross section for $W^{+}$ 
or $W^{-}$ boson production.  
The parton distribution functions (PDFs) describing the internal 
structure of the proton are constrained by measuring $A(\yW)$~\cite{cskim}.

Previous measurements~\cite{run1, run2, d0murun2, d0elerun2} of the $W$ charge
asymmetry at the Tevatron were made as a function of the
pseudorapidity $\eta$~\cite{coord} of the leptons from decays of 
$W \rightarrow l \nu_{l} ~(l = e, \mu)$ since
the $W$ decay involves a neutrino whose longitudinal momentum is not
determined experimentally. However, the lepton charge asymmetry is a
convolution of the $W$ production charge asymmetry and the $V-A$
asymmetry from $W$ decays. These two asymmetries tend to cancel at
large pseudorapidities ($|\eta| \gtrsim 2.0$), and the convolution weakens and
complicates the constraint on the proton PDFs.  

In the measurement presented in this Letter, the complication is
resolved by using additional information in the lepton transverse
energy ($\et$) and the missing transverse energy ($\met$)
~\cite{coord} on an event-by-event basis to measure the asymmetry as a
function of the $|\yW|$ instead of the lepton $|\eta|$.  This new 
analysis technique~\cite{wpca-prdrc} gives
the first direct measurement of the $W$ production charge asymmetry
using $\wenu$ decays.  We use data from 1~$\ifb$ of
integrated luminosity collected by the CDF II detector.
The region of acceptance is $|\yW| < 3.0$, giving the new measurement
an ability to improve proton PDFs determinations for $0.002 \lesssim x
\lesssim 0.8$, where $x$ is the fraction of the proton momentum
carried by $u$- or $d$-type quarks.  This analysis is described in
detail in~\cite{thesis}.

The CDF II detector is described in detail elsewhere
~\cite{cdf-det}. What follows is a brief description of the detector
components needed to identify $\wenu$ events, which are characterized
by large missing transverse energy ($\met$) and a track in the central
drift chamber (COT)~\cite{cdf-cot} or in the silicon tracking system
(SVX)~\cite{cdf-svx,cdf-isl} that points to a cluster of energy in the
electromagnetic (EM) calorimeters ~\cite{cdf-cemcal,cdf-pemcal}.  The
SVX provides precise track measurements from eight radial layers of
microstrip sensors.  The COT provides additional tracking information
from 96 layers of wires. Tracks are measured inside a 1.4 T solenoidal
magnetic field that allows electron charge determination from the
curvature of the track. The COT allows track reconstruction in the
range $|\eta| \lesssim 1.6$, while the SVX extends the capability up
to $|\eta| \simeq 2.8$.  Outside the tracking system, EM and hadronic
(HAD) calorimeters measure the energies of showering particles.  The
calorimeters are divided into two types: a central calorimeter with a
fiducial region covering $|\eta| < 1.1$, and a forward calorimeter
covering $1.2 < |\eta| < 3.5$.

We use two types of $\wenu$ events, classified by the calorimeter
section in which the electron is detected.  The data are initially
selected by an on-line event selection (trigger) system. 
The trigger for the central electrons requires an EM energy cluster 
with $\et > 18~ \GeV$ and a matching track with $\pt > 9~ \GeV$. 
The forward trigger, designed specifically for
$W$ candidates, requires an EM energy cluster with $\et > 20~ \GeV$
and $\met > 15 ~\GeV$.  

For central electrons, we require off-line event selection including
an isolated energy cluster in the region $|\eta| < 1.1$ with $\et >
25~ \GeV$ and Iso$(0.4) < 4.0$ GeV.  The isolation Iso$(0.4)$ is
defined as the calorimeter energy contained within a cone of radius $R
= \sqrt{(\Delta\eta)^2 + (\Delta\phi)^2} = 0.4$~\cite{coord} around
the electron direction excluding the energy associated with the
electron.  A more detailed description of the central electron
selection can be found in~\cite{wzxsec}. The forward electrons are
selected by requiring an isolated energy cluster with $\et > 20~
\GeV$, the ratio of energy detected in the HAD and EM calorimeters to
be less than 0.05.  The tracks are reconstructed using COT information
in the region $|\eta| <1.6$, while at higher $|\eta|$ tracks are
reconstructed using the SVX detectors alone.  In order to reduce the
charge misidentification and backgrounds, additional
requirements for the forward tracks are imposed such as requiring
the extrapolated charged-particle position to be consistent with the
position measured in the calorimeter.  Candidate $\wenu$ events are
required to have exactly one $e^{\pm}$ as well as $\met >
25~ \GeV$. The final $\wenu$ data sample contains 537,858 events with
a central electron and 176,941 forward electron events. To evaluate
the detector acceptance and resolution for $\wenu$ events we use the
\textsc{pythia}~\cite{pythia} event generator followed by the CDF
detector simulation.

We determine the neutrino's longitudinal momentum, within a two-fold
ambiguity, by constraining the $e \nu$ mass to be that of the $W$ boson. 
This ambiguity can be resolved on a statistical basis from the
known $V-A$ decay distribution using the decay angle between the
electron and the proton in the $W$ rest frame, $\theta^{*}$, and from
the $W^{+}$ and $W^{-}$ production cross sections as a function of $W$
rapidity ($d\sigma^{\pm}/dy_{W}$). To do this we assign a weighting
factor to the two rapidity solutions, depending on the charge of the
$W$ boson, $w^{\pm}_{1,2}$:
\begin{equation}
w^{\pm}_{1,2} = \frac{P_{\pm}(\cos\theta^*_{1,2},y_{1,2},p_T^W) \sigma^{\pm}(y_{1,2})}{P_
{\pm}(\cos\theta^*_{1},y_{1},p_T^W) \sigma^{\pm}(y_{1}) + P_{\pm}(\cos\theta^*_{2},y_{2},p_T
^W) \sigma^{\pm}(y_{2})},
\label{eq:wt} 
\end{equation}
where
\begin{equation}
P_{\pm}(cos\theta^*,y_W,p_{T}^{W}) = (1 \mp cos\theta^*)^2 + Q(y_W,p_{T}^{W})(1 \pm cos\theta^*)^2. \label{eq:prob}
\end{equation}

The $\pm$ signs indicate the $W$ charge and the indices 1 and 2 are
for the two $W$ rapidity solutions.  The differential cross section as
a function of $y_{W}$ is determined using a next-to-next-to-leading
order (NNLO) QCD calculation~\cite{nnlo} using the MRST 2006 NNLO
PDFs~\cite{mrst2006}.  The ratio of the $(1 + \cos\theta^*)^2$ to $(1
- \cos\theta^*)^2$ angular distributions, $Q(y_W,p_{T}^{W})$, in
Eq.~\ref{eq:prob} is determined by the quark versus anti-quark
composition of the proton using the event generator
\textsc{mc@nlo}~\cite{mcatnlo}.  This ratio is evaluated as a function
of $y_{W}$ and the $W$ transverse momentum $p_{T}^{W}$.  Although the
weighting factor given by Eq.~\ref{eq:wt} depends primarily on the
$W^+$ and $W^-$ cross sections, it does have some weak dependence on
the input $W$ charge asymmetry. Therefore, this method requires us to
iterate the procedure to eliminate this dependence.

Correct charge identification is crucial for the measurement of the
charge asymmetry measurement, because it directly affects the yield
for a particular charge and $\yW$ and is corrected for on an
event-by-event basis.  The charge misidentification rate (Charge
MisID) is measured as a function of $\eta$ using $\zeenc$ events where
both electrons are identified as having the same charge sign.  The
misidentification rate ranges from $(0.18 ~\pm~ 0.05)\%$ for $|\eta| <
1.1$ to $(17.26 ~\pm~ 2.02)\%$ for $|\eta| > 2.04$.

The $A(y_W)$ values are corrected for the backgrounds to $\wenu$
candidates. We consider $\wtnu$, where the $\tau$ decays leptonically
to an electron plus neutrinos, as contributing to the signal and is
included in the overall signal acceptance. The background fractions
due to $\zee$ events where one of the electrons is not reconstructed
and mimics a neutrino are (0.59 $\pm$ 0.02)$\%$ for central electrons
and (0.54 $\pm$ 0.03)$\%$ for forward electrons.  The small
contamination from the $\ztau$ process is found to be (0.10 $\pm$
0.01)$\%$ for both central and forward electrons.  The background from
misidentified jets (QCD) is estimated by fitting the isolation
distribution of electrons.  Electrons in the calorimeter are
characterized by having most of their energy deposited within an
isolation cone centered on the electron, while jets may have
significant energy deposits outside this cone.  The QCD background
fraction for central and forward electrons are (1.21 $\pm$ 0.21)$\%$
and (0.67 $\pm$ 0.18)$\%$, respectively.

\begin{table*}
\caption{Statistical and systematic uncertainties for the $W$ production charge asymmetry. All values are ($\times 10^{-2}$) and show the correlated uncertainties for both positive and negative rapidities.
}
\begin{ruledtabular}
\begin{center}
\begin{tabular}{c c c c c c c c c}
$|y_{W}|$  & Charge & Back- & Energy Scale      & Recoil & Electron & Electron       & PDFs & Stat. \\
          & MisID & grounds& $\&$ Resolution   & Model  & Trigger  & ID             &     &           \\ \hline
0.0 - 0.2 & 0.02 & 0.04 & 0.01 & 0.11 & 0.03 & 0.02 & 0.03 & 0.31 \\
0.2 - 0.4 & 0.01 & 0.09 & 0.04 & 0.22 & 0.08 & 0.07 & 0.08 & 0.32 \\
0.4 - 0.6 & 0.02 & 0.11 & 0.06 & 0.22 & 0.13 & 0.17 & 0.15 & 0.33 \\
0.6 - 0.8 & 0.03 & 0.15 & 0.07 & 0.34 & 0.14 & 0.30 & 0.22 & 0.32 \\
0.8 - 1.0 & 0.03 & 0.20 & 0.07 & 0.42 & 0.11 & 0.47 & 0.24 & 0.34 \\
1.0 - 1.2 & 0.04 & 0.18 & 0.08 & 0.33 & 0.09 & 0.69 & 0.27 & 0.38 \\
1.2 - 1.4 & 0.05 & 0.18 & 0.15 & 0.67 & 0.06 & 0.78 & 0.28 & 0.43 \\
1.4 - 1.6 & 0.04 & 0.14 & 0.14 & 1.10 & 0.04 & 0.85 & 0.28 & 0.50 \\
1.6 - 1.8 & 0.08 & 0.12 & 0.26 & 0.92 & 0.03 & 0.89 & 0.29 & 0.55 \\
1.8 - 2.05 & 0.22 & 0.13 & 0.31 & 0.82 & 0.06 & 0.80 & 0.34 & 0.62 \\
2.05 - 2.3 & 0.44 & 0.21 & 0.53 & 0.59 & 0.17 & 0.85 & 0.42 & 0.83 \\
2.3 - 2.6 & 0.45 & 0.19 & 0.62 & 0.40 & 0.27 & 0.86 & 0.50 & 1.10 \\
2.6 - 3.0 & 0.14 & 0.10 & 0.60 & 0.43 & 0.28 & 0.65 & 0.53 & 2.30
\end{tabular}
\label{tab:syst_err}
\end{center}
\end{ruledtabular}
\end{table*}

The scale and resolution of the electromagnetic calorimeter energy and
the missing transverse energy can affect the measured $W$
rapidity and thus the asymmetry measurement.  The EM calorimeter
energy scale and resolution are tuned in the simulation to reproduce
the $\zee$ mass peak in data.  The uncertainties on the
energy scale and resolution for central electrons are measured to be
$\pm~ 0.05\%$ and $\pm~ 0.07\%$; for forward electrons they are $\pm~
0.3\%$ and $\pm~ 0.8\%$, respectively.  
The hadronic showering, the boson recoil-energy, and the
underlying event energy of the hadronic calorimeter energy measurement 
play important roles in determining the \met. 
The simulation for the calorimeter deposition in $\wenu$ events is tuned 
to provide the best possible match with data, including its dependence on 
$\eta$. The uncertainty on the transverse recoil energy scale is 
$\pm~ 0.3\%$ and $\pm~ 1.4\%$ for central and forward electrons, 
respectively.

We also investigate potential sources of a charge bias and $\eta$
dependence in the kinematic and geometrical acceptance of the event
(estimated with simulated data) and efficiencies of the trigger and
the electron identification (measured with data).  The trigger
efficiencies for the central and forward electrons are measured using
data from independent triggers. We find the trigger efficiencies do
not depend on charge, but do depend on the $\eta$ and $\et$ of the
electron. The average trigger efficiencies for the central and forward
electrons are $(96.1 ~\pm~ 1.0)\%$ and $(92.5 ~\pm~ 0.3)\%$,
respectively.  Electron identification and track matching efficiencies
(ID) are measured in data and simulation using the $\zee$ channel.

The choice of PDF sets has an effect on the shape of the
$d\sigma^{\pm}/dy_{W}$ distribution, as well as on the ratio of quarks
and anti-quarks in the angular decay distribution. We use the 40
CTEQ6.1 error PDF sets~\cite{cteq6m} and re-determine the
$d\sigma^{\pm}/dy_{W}$ production cross section and the angular
distribution of $\cos\theta^{*}$ for each error PDF set.

As expected, the data are found to be invariant under CP
transformations $A(\yW) = -A(-\yW)$, the two sets of points are in
statistical agreement, so we combine the $A(\yW)$ bins with the
complementary $-A(-\yW)$ bins in order to improve the precision of
this measurement. We quote the statistical combination of $A(\yW)$
with $-A(-\yW)$, using the Best Linear Unbiased Estimate (BLUE)
method~\cite{blue}, accounting for all correlations for both positive
and negative $y_W$ bins. The statistical correlation coefficient
between bins is found to be $< 0.05$.  Table~\ref{tab:syst_err}
summarizes the statistical and systematic uncertainties on $A(|y_W|)$.

The measured asymmetry $A(|y_W|)$, which combines the positive and
negative $y_W$ bins, is shown in Fig.~\ref{fig:asym_fold}. Also shown
are the predictions of a NNLO QCD calculation using the MRST 2006 NNLO
PDF sets and a NLO QCD calculation using the CTEQ6.1 NLO PDF sets,
which are in agreement with the measured asymmetry.  Values of
$A(y_W)$ and the total uncertainty for each $|y_{W}|$ bin are listed
in Table~\ref{tab:wasym}.  Since this measurement depends on the width
of the $W$, in particular for the highest $y_W$ bin, the bin centers
account for the $W$ rapidity and $W$ mass range accepted in each bin.
We correct the bin centers to the value of $<|y_W|>$ (average of
$W^{+}$ and $W^{-}$ rapidities) for which the asymmetry is equal to
the one for a fixed $W$ mass of $80.403 ~\rm{GeV/c^{2}}$.
\begin{figure}
  \begin{center}
    \includegraphics[width=\figsize]{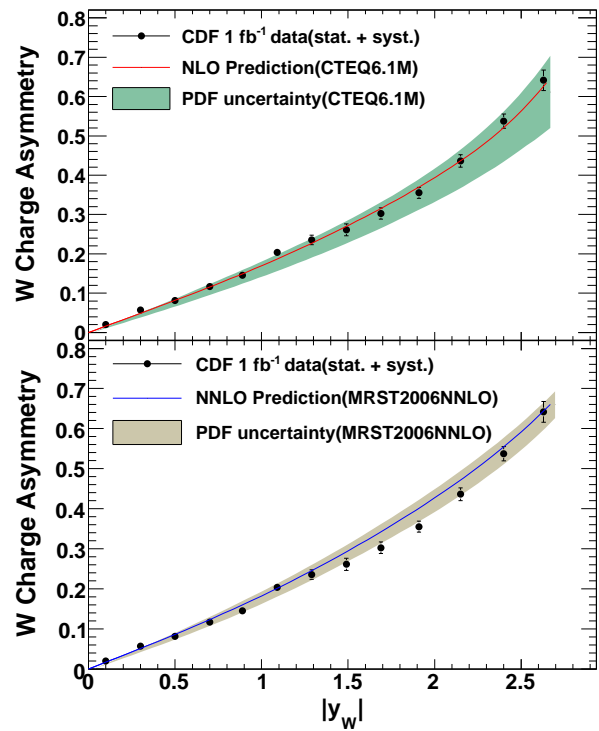}
    \caption{The measured $W$ production charge asymmetry and predictions from 
	(a) NLO CTEQ6.1  and 
	(b) NNLO MRTST2006, with their associated PDF uncertainties.}
  \label{fig:asym_fold}
  \end{center}
\end{figure}
\begin{table}
\caption{The $W$ production charge asymmetry with total systematic and statistical uncertainties.}
\begin{ruledtabular}
\begin{center}
\begin{tabular}{c c c c c c }
    $|y_{W}|$  & $<|y_{W}|>$ & $A(y_{W})$ & $\sigma_{sys}$ & $\sigma_{sys+stat}$\\ \hline
0.0 - 0.2 & 0.10 & 0.020 & $\pm$0.001 & $\pm$0.003 \\
0.2 - 0.4 & 0.30 & 0.057 & $\pm$0.003 & $\pm$0.004 \\
0.4 - 0.6 & 0.50 & 0.081 & $\pm$0.004 & $\pm$0.005 \\
0.6 - 0.8 & 0.70 & 0.117 & $\pm$0.006 & $\pm$0.006 \\
0.8 - 1.0 & 0.89 & 0.146 & $\pm$0.007 & $\pm$0.008 \\
1.0 - 1.2 & 1.09 & 0.204 & $\pm$0.008 & $\pm$0.010 \\
1.2 - 1.4 & 1.29 & 0.235 & $\pm$0.011 & $\pm$0.012 \\
1.4 - 1.6 & 1.49 & 0.261 & $\pm$0.014 & $\pm$0.015 \\
1.6 - 1.8 & 1.69 & 0.303 & $\pm$0.014 & $\pm$0.014 \\
1.8 - 2.05 & 1.91 & 0.355 & $\pm$0.013 & $\pm$0.014 \\
2.05 - 2.3 & 2.15 & 0.436 & $\pm$0.013 & $\pm$0.016 \\
2.3 - 2.6 & 2.40 & 0.537 & $\pm$0.014 & $\pm$0.018 \\
2.6 - 3.0 & 2.63 & 0.642 & $\pm$0.012 & $\pm$0.026 
\end{tabular}
\label{tab:wasym}
\end{center}
\end{ruledtabular}
\end{table}

In conclusion, using a new analysis technique we report the first 
direct measurement of the $W$ boson charge asymmetry from Run II 
of the Tevatron, using data from 1 $\ifb$ of integrated luminosity 
taken with the CDF II detector. 
Since the total uncertainties are smaller 
than the uncertainties coming from PDFs, as is also shown 
in~\cite{wpca-prdrc}, this direct measurement of the asymmetry is 
more sensitive to the ratio of d/u momentum
distributions in the proton at high $x$ than previous lepton 
charge asymmetry measurements.
This result is therefore expected to improve the precision of the global
PDFs fits.

We wish to thank R.S. Thorne and W.J. Stirling for useful discussions
on the theoretical predictions.  We thank the Fermilab staff and the
technical staffs of the participating institutions for their vital
contributions. This work was supported by the U.S. Department of
Energy and National Science Foundation; the Italian Istituto Nazionale
di Fisica Nucleare; the Ministry of Education, Culture, Sports,
Science and Technology of Japan; the Natural Sciences and Engineering
Research Council of Canada; the National Science Council of the
Republic of China; the Swiss National Science Foundation; the
A.P. Sloan Foundation; the Bundesministerium f\"ur Bildung und
Forschung, Germany; the Korean Science and Engineering Foundation and
the Korean Research Foundation; the Science and Technology Facilities
Council and the Royal Society, UK; the Institut National de Physique
Nucleaire et Physique des Particules/CNRS; the Russian Foundation for
Basic Research; the Ministerio de Ciencia e Innovaci\'{o}n, and
Programa Consolider-Ingenio 2010, Spain; the Slovak R\&D Agency; and
the Academy of Finland.

\end{document}
%